\documentclass[%
 reprint,
 amsmath,amssymb,
 aps,
]{revtex4-1}
\usepackage{tikz}
\usepackage{graphicx}
\usepackage{pgfplots}
\usepackage{dcolumn}
\usepackage{bm}
\usepackage{subfigure}
\usepackage{multirow}
\usetikzlibrary{plotmarks}
\usepackage{lipsum}
\usepackage{tikz}
\usepackage{transparent}

\begin{document}

\preprint{APS/123-QED}

\title{Surface flow profiles for dry and wet granular materials by Particle Tracking Velocimetry; the effect of wall roughness}

\author{Sudeshna Roy}
\email{s.roy@utwente.nl}
\author{Bert J. Scheper}%
\author{Harmen Polman}%
\author{Anthony R. Thornton}%
\author{Deepak R. Tunuguntla}%
\author{Stefan Luding}%
\author{Thomas Weinhart}%

\affiliation{
 Multi-Scale Mechanics, Engineering Technology (ET) and MESA+ \\
 University of Twente, 7500 AE Enschede, The Netherlands\\
}%

\date{\today}

\begin{abstract}
Two-dimensional Particle Tracking Velocimetry (PTV) is a promising technique to study the behaviour of granular flows. The aim is to experimentally determine the free surface width and position of the shear band from the velocity profile to validate simulations in a split-bottom shear cell geometry. The position and velocities of scattered tracer particles are tracked as they move with the bulk flow by analyzing images. We then use a new technique to extract the continuum velocity field, applying coarse-graining with the postprocessing toolbox MercuryCG on the discrete experimental PTV data. For intermediate filling heights, the dependence of the shear (or angular) velocity on the radial coordinate at the free surface is well fitted by an error function. From the error function, we get the width and the centre position of the shear band. We investigate the dependence of these shear band properties on filling height and rotation frequencies of the shear cell for dry glass beads for rough and smooth wall surfaces. For rough surfaces, the data agrees with the existing experimental results and theoretical scaling predictions. For smooth surfaces, particle-wall slippage is significant and the data deviates from the predictions. We further study the effect of cohesion on the shear band properties by using small amount of silicon oil and glycerol as interstitial liquids with the glass beads. While silicon oil does not lead to big changes, glycerol changes the shear band properties considerably. The shear band gets wider and is situated further inward with increasing liquid saturation, due to the correspondingly increasing trend of particles to stick together.
\end{abstract}

\pacs{Valid PACS appear here}
\maketitle

\section{Introduction}
Dense granular materials display rich and complex flow properties, which differentiate them from ordinary fluids. However, experimental studies of granular flows are difficult to perform due to their opaque nature. Nevertheless, insight has been generated using Magnetic Resonance Imaging \cite{kawaguchi2010mri,hill1997bulk,nakagawa1993non,ehrichs1995granular}, X-ray imaging \cite{baxter1989pattern}, radioactive tracers \cite{harwood1977powder}, freezing granules in resin \cite{ratkai1976particle}, digital imaging \cite{guler1999measurement,capart2002voronoi,bonamy2002experimental,lueptow2000piv}, Particle Image Velocimetry (PIV) applied to quasi-2D granular flows \cite{bokkers2004mixing,laverman2008investigation,zeilstra2008experimental,jarray2019wet}, and Particle Tracking Velocimetry (PTV) \cite{chou2009cross,yang2006effect,liao2009influence,shirsath2015cross,sokoray2008particle}. PTV is a method of analysis which relies on the ability to track the evolution of position of individual tracer particles which is an attractive characteristic for the study of granular flows. In the last decades, efforts have been made to improve PTV with new algorithms, e.g. their reliability at high densities of the tracer particles \cite{nitsche2009imaging,lei2012vision,jiang2017improved}. Therefore, PTV has become the method of choice to analyse measurements from moving particle flows, e.g., in a shear cell.
\par
Until recently, it was mostly reported that all granular shear bands are narrow, i.e. a few particle diameters wide and accompanied
by strong localisation of strain \cite{veje1999kinematics,utter2004self,jasti2008experimental}. However, in 2003 Fenistein \textit{et al.} \cite{fenistein2003kinematics} discovered that in a modified Couette cell, granular shear bands can be arbitrarily broad (see Figure \ref{Setup}). In this geometry, particles are confined under gravity in the annular space between a fixed inner cylinder and a rotating outer cylinder. The particles remain stationary near the inner cylinder and rotate like a solid near the outer cylinder; a shear band separates the two regions (represented by the shaded region in Figure \ref{Setup}(b)). For very shallow packing, the shear band measured at the top surface is narrow and located at $r = R_\mathrm{s}$, see Figure \ref{Setup}(b). As the filling height of the material $H$ increases, the shear band width increases in radial direction, moving towards the inner cylinder. For sufficiently large $H$, the shear band overlaps with the inner cylinder at $r = R_\mathrm{in}$. Unger \textit{et al.} \cite{unger2004shear} predicted that the shape of the boundary, between moving and stationary material, would undergo a first-order transition as $H$ is increased beyond a threshold value $H^*$. The shearing region for $H < H^*$ is open at the top, but intersects the free surface and abruptly collapses to a closed cupola completely buried inside the bulk for $H > H^*$. Note that our focus in this paper is on the region $H < H^*$ where the shearing region lies exposed at the free surface. Here, the position of the shear band centre can be predicted as a function of height in the system based on the \textit{principle of least dissipation of energy}. Ries \textit{et al.} \cite{ries2007shear} measured the width of the shear band in the bulk as a function of height in the bulk of the system from DPM simulations. 
\par
Previous studies focused primarily on the surface and bulk flows in shallow modified Couette cells for dry granular flows \cite{dijksman2009granular,dijksman2010granular,cheng2006threedimensional}. The question arises regarding how the presence of attractive forces affects shear banding. So far, only a few attempts have been made to answer this question concerning dense metallic glasses \cite{spaepen1977microscopic,li2002nanometre}, adhesive emulsions \cite{becu2006yielding,chaudhuri2012inhomogeneous}, attractive
colloids\cite{vermant2001large,hohler2005rheology,coussot2010physical}, cemented granular media \cite{estrada2010simulation},  wet granular media \cite{mani2012fluid,schwarze2013rheology}, and clayey soils \cite{yuan2013experimental}. Singh \textit{et al.} \cite{singh2014effect} studied the effect of dry cohesion on the width and the position of the shear band from DPM simulations, including a detailed report on the local rheology and micro-structure. Here, the dry cohesion is varied by varying the strength of the cohesive force. The effect of cohesion on the shear band properties depend on the Bond number $Bo$, which is the strength of the cohesive force relative to the confining force. The general conclusion was, both, the position and the width of the shear band remain unaffected in a cohesive system for $Bo < 1$, while the shear band moves inward and becomes wider for $Bo >= 1$. In case of wet granular materials, the liquid bridge interactions between particles are influenced by long range forces which should have additional effect on the bulk behaviour. Thus, in the present study, we intend to experimentally investigate the surface flow profile of, both, dry and wet granular materials. 

\par
Continuum fields often have to be extracted from discrete particle data to validate and analyse the behaviour of stationary or transient particulate system. One such approach is by applying Coarse Graining (CG) technique. This method has several advantages: (i) the fields automatically satisfy the conservation equations of continuum mechanics (ii) particles are not assumed to be rigid or spherical and (iii) the results are valid for single particles (no averaging over ensembles of particles). This technique was put forward by \cite{goldhirsch2010stress} and has been applied extensively to obtain continuum fields from discrete particle simulations \cite{weinhart2012discrete,thornton2012modeling,weinhart2012closure,weinhart2013coarse,tunuguntla2016discrete,tunuguntla2017discrete,tunuguntla2017comparing}. In our study, we apply the coarse graining technique to the discrete particle data obtained from the experiments. In our experiments, we use a mix of transparent bulk and coloured tracer particles to explore the technique of Particle Tracking Velocimetry for measuring the velocity of the seeded tracer particles at the flow's surface. Thereby, we obtain discrete particle data from the experiments. We follow a novel approach and apply the aforementioned CG technique to the discrete particle experimental data to obtain continuum velocity fields using MercuryCG toolbox, which is a part of the open source code MercuryDPM \cite{weinhart2016mercurydpm,weinhart2017mercurydpm,thornton2013review,roy2016micro,roy2017general,roy2018liquid}. Moreover, our exploration in this paper is not limited to dry granular materials; we apply the novel approach of PTV-CG to obtain velocity fields of wet granular materials as well. The primary challenges are to track the tracer particles (i) in a wet-particle system and (ii) in a system where particles are closely packed.
\par
We present the experimental set-up in Sec. \ref{setup}, an experimental overview in Sec. \ref{overview}, the PTV methodology in Sec. \ref{PTV} and the CG technique for experimental data in \ref{coarse}. The new application of the combination of the PTV methodology and the CG technique is termed as PTV-CG in this paper. The observations for dry and wet glass beads experiments are reported in Sec. \ref{dry} and \ref{wet}, respectively. Finally, we conclude in Sec. \ref{con}.

\section{Experimental set-up}\label{setup}
We probe the rheology of granular media in a split-bottom shear cell shown in Figure \ref{Setup}. The geometry, proposed by Fenistein \textit{et al.} \cite{fenistein2003kinematics}, is a modification of the Couette cell. This set-up allows the formation of a wide shear band near the free surface, free from the boundary effects. 

\subsection{Geometry}

\begin{figure}
\center
\begin{tikzpicture}
\node[anchor=south west,inner sep=0] at (-1,5) {\includegraphics[width=0.7\columnwidth]{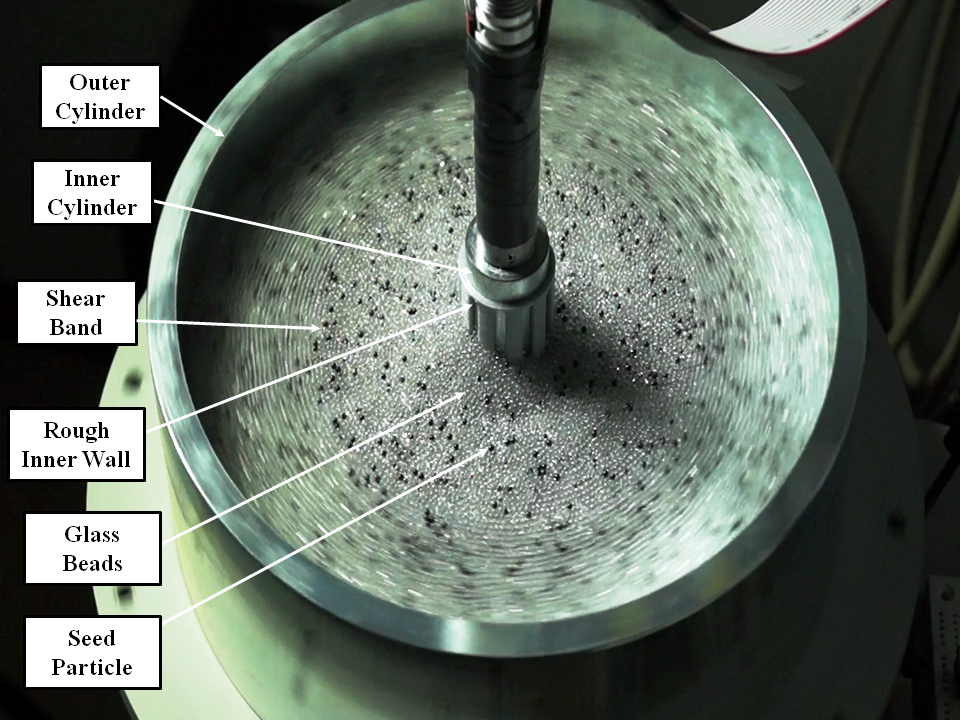}};

\node[anchor=south west,inner sep=0] at (-1,1) {\includegraphics[width=0.7\columnwidth]{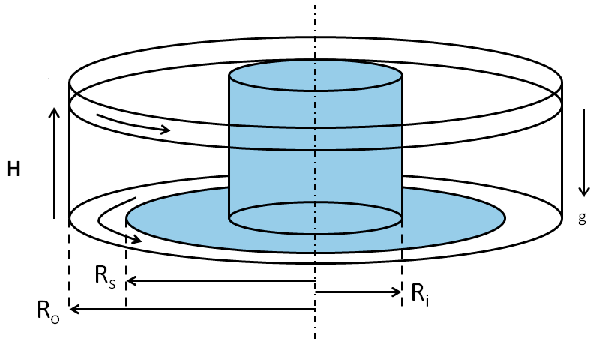}};
     \shade[color=red] (4,2.5) parabola (3.3,3.15) -- (3.8,3.2)(4,2.5) parabola (3.8,3.2);

    \draw (4,2.5) parabola (3.3,3.15);
  \draw (4,2.5) parabola (3.8,3.2);
\end{tikzpicture} 

 \caption{The experimental setup of the split-bottom shear cell (top) and schematic showing the set-up and the shaded region as the shear band (bottom).}
    \label{Setup}
\end{figure}
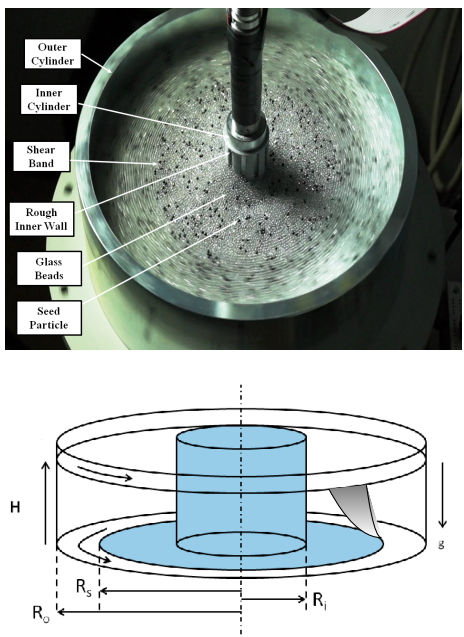

The shear cell consists of a rotating outer cylinder of radius $R_\mathrm{o} = 110$ mm, a stationary inner cylinder of radius $R_\mathrm{i} = 14.7$  mm and a split at the bottom of the shear cell at $R_\mathrm{s} = 83$ mm. This split separates the rotating outer part from the stationary inner part of the shear cell. The annular space between the inner and the outer cylinder is filled with glass beads upto a given filling height $H$. The outer cylinder rotates at a constant frequency $f_\mathrm{rot}$. The rotational motion is driven by a motor attached to the bottom of the setup. We do experiments with dry glass beads for varying filling heights $H$ and rotation frequencies $f_\mathrm{rot}$ as mentioned in Tables \ref{appH} and \ref{appF}, respectively. We do experiments with wet glass beads, varying the liquid content, keeping the filling height $H$ and the rotation frequency $f_\mathrm{rot}$ constant. Figure \ref{Setup} (top) shows the experimental set-up, as captured by the high speed camera. The moving region is indicated by the blurry particles and the static region indicated by the sharply defined particles. A schematic figure for the same set-up is also shown in Figure \ref{Setup} (bottom). 

\subsection{Particles} \label{size}
We use transparent glass beads from Sigmund Lindner, SiLi beads Type S and opaque black particles of same specifications as tracers. According to the size specification provided by the manufacturer, the particles have a normal size distribution with mean diameter $d_\mathrm{p} = 1.70$ mm and standard deviation $\pm 0.083$ mm. The density of Type S particles is $2500$ kgm$^{-3}$. We measure the velocities of the tracer particles by image analysis of the surface snapshots. 
\subsection{Wall roughness}
\begin{figure}[!htb]
  \begin{center}
  {%
  \includegraphics[width=0.7\columnwidth]{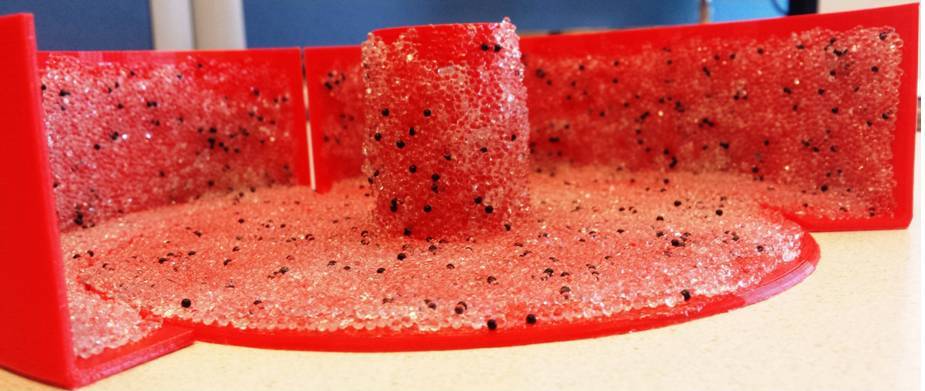}
  \hfill
	\includegraphics[width=0.7\columnwidth]{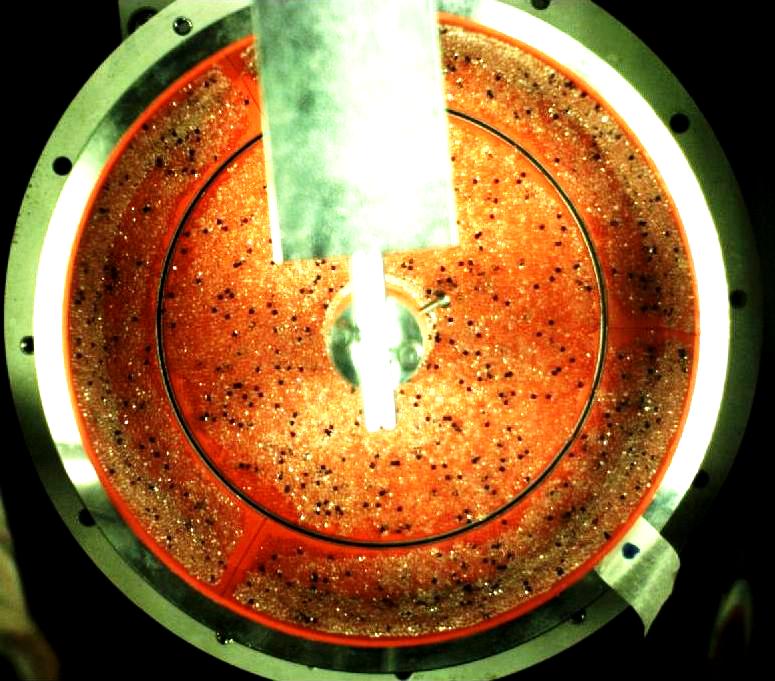}

   }%
  \end{center}
  \caption{Side view (top) and top view (bottom) of the rough wall insert, showing particles glued on the surface of the insert.}
  \label{Wall}
\end{figure}

Particle wall friction has a major influence on the bulk behaviour of granular materials. In order to understand the effect of wall friction, we perform experiments with smooth and rough walls. This is done by gluing particles, identical to those used for the experiments, on the walls and bottom surface of a red colour 3D-printed insert as shown in Figure \ref{Wall} (top). Note that we glue double layer of particles so that there is no gap left free on the surfaces. The insert is fastened inside the shear cell as shown in Figure \ref{Wall} (bottom). Thus, the roughness scale is equivalent to a particle diameter. Alternatively, we performed experiments for dry granular materials with smooth walls, i.e. by removing the insert of glued particles, see Secs. \ref{Cases} and \ref{rot}. All other experiments are done with rough walls as shown in the set-up in Figure \ref{Wall}.
\subsection{Liquids and concentrations}

Our primary focus in the experiments was to determine the shear band properties from the surface velocity profile, obtained by 2D PTV. We do preliminary studies on the properties of dry glass beads. In the later section of this paper, we discuss the effect of adding aqueous glycerol as an
interstitial liquid on the bulk flow properties. We measured the contact angle properties of glycerol on the glass bead surface by Sessile drop test. Aqueous glycerol solution ($80\%$ glycerol) has an average contact angle of $24^\circ$ on the glass surface. Aqueous glycerol solutions has a surface tension of $0.064$ Nm$^{-1}$ at standard temperature \cite{surf}. 
\par
We vary the bulk saturation of particles with glycerol in a set of experiments. The bulk saturation of liquid $S^*$ in the system is defined as the ratio of the liquid volume to the pore volume \cite{weigert1999calculation}. The liquid saturation of glycerol is varied according to Table \ref{appG}. The given saturation are within the pendular regime of liquid bridge ($S^* \approx 0.30$) where the effects of capillary bridges are most dominant. 
\section{Overview of the Experiments}\label{overview}

\begin{table*}[htb!]
\caption{Abbreviations and description for the experiments.}
\label{abb}       
\begin{tabular}{llll}
\hline\noalign{\smallskip}
\textbf{Abbreviations} & \multicolumn{3}{c}{\textbf{Description}}\\ 
\hline\noalign{\smallskip}
\textbf{S} & & & Experiments with Smooth Walls \\
\textbf{R} & & & Experiments with Rough Walls \\
\textbf{h} ($f_\mathrm{rot} = 0.03$ s$^{-1}$) & & & Experiments of dry glass beads with varying filling heights \\
\textbf{f} ($H = 36$ mm) & & & Experiments of dry glass beads with varying rotation rates \\
\textbf{g} ($H = 36$ mm, $f_\mathrm{rot} = 0.03$ s$^{-1}$) & & & Experiments of wet glass beads with varying saturation of liquid Glycerol\\ 
\textbf{o} ($H = 36$ mm, $f_\mathrm{rot} = 0.03$ s$^{-1}$) & & & Experiments of wet glass beads with varying saturation of liquid Silicon Oil\\ 
\noalign{\smallskip}\hline
\end{tabular}
\end{table*}
 \begin{table*}[htb!]
\caption{List of the experiments nomenclature and the respective parameters. }
\label{des}       
\begin{tabular}{lllllllllllllll}
\hline\noalign{\smallskip}

{\textbf{Abbreviations}} & \multicolumn{12}{c}{\textbf{Parameters}}\\ 
\hline\noalign{\smallskip}
\textbf{Sh} ($H~ [\mathrm{mm}]$) &  & 13 & 18 & 23 & 28 & 32 & 38 &  &  &  &  &  &\\ 

\textbf{Rh} ($H~ [\mathrm{mm}]$) &  & 6 & 8 & 10 & 13 & 15 & 18 & 20 & 23 & 25 & 28 & 32 & 38\\ 

\textbf{Sf} ($f~ [\mathrm{s}^{-1}]$) &  & 0.01 & 0.03 & 0.07 & 0.19 & 0.50 &  &  &  &  &  &  &\\ 

\textbf{Rf} ($f~ [\mathrm{s}^{-1}]$) &  & 0.01 & 0.03 & 0.07 & 0.19 & 0.50 &  &  &  &  &  &  &\\ 

\textbf{Sg} ($S^*~ [~]$) &  & 0 & 0.08 & 0.17 & 0.25 & 0.41 &  &  &  &  &  &  &\\ 

\textbf{Rg} ($S^*~ [~]$) &  & 0 & 0.002 & 0.004 & 0.009 & 0.013 & 0.017 & 0.022 & 0.044 & 0.087 & 0.131 & 0.175 & 0.218\\ 

\textbf{So} ($S^*~ [~]$) &  & 0 & 0.08 & 0.17 & 0.25 & 0.41 &  &  &  &  &  &  &\\ 

\noalign{\smallskip}\hline

\end{tabular}
\end{table*}
In this section, we elaborate the list of the experiments that are performed. We do different experiments and denote them through combination of the abbreviations given in Table \ref{abb}. The first letter of our nomenclature denotes the kind of wall surfaces \textbf{S} for smooth walls and \textbf{R} for rough walls. The second letter denotes the varying parameter for the given set of experiments as described in Table \ref{abb}. For example, an abbreviation $\bf{Sh}$ denotes experiments done with varying filling height $H$ with dry glass beads at a given rotation frequency. $\bf{Sg}$ denotes experiments done with smooth walls with glass beads and varying saturation $S^*$ of the interstitial liquid glycerol at a given rotation rate and filling height. For more details, see Table \ref{des}. Note that all the experiments with fixed rotation rate and fixed filling height are done with rotation frequency $f_\mathrm{rot} = 0.03$ s$^{-1}$ and filling height $H = 36$ mm, respectively. Table \ref{des} describes the varying range of parameters for each set of experiments. The experiments corresponding to the smooth walls  of the shear cell were done in 2016 and those with the rough walls were done recently in 2018. Hence, the ranges and number of data for different experiments vary. Results from experiments in smooth wall (\textbf{S}) shear cell with glass beads and interstitial liquid as glycerol and silicon oil, represented as \textbf{Sg} and \textbf{So} respectively, are referred to in \cite{roy2018hydrodynamic} and are not shown in this paper. We do two independent consecutive rounds of experiments for each data set, denoted as suffix $\bf{1}$ and $\bf{2}$ respectively, in the results sections. However, only one set of experiments is done for the data corresponding to \textbf{Sf}.
\par
We also did experiments with varied saturation of Silicon oil as an interstitial liquid. The results are shown in \cite{roy2018hydrodynamic}. However, the use of silicon oil is found to have insignificant role in changing the shear band properties as compared to glycerol. Although, the width changes by approximately $13$\%, the shear band centre position shifts by only $2$\%. Hence, we focus on the role of interstitial liquid glycerol only in this paper.
\section{Velocity measurement}
Granular flows are often inhomogeneous in the presence of a shear band. In geometries such as inclined-plane flows, avalanches and Couette flows, shear bands are narrow \cite{schall2009shear}. To study wide shear bands, we use a modified split-bottom shear cell where the granular flow is driven from the bottom, instead of from the side walls as in Couette flows \cite{fenistein2004universal}. The differential motion of the outer and the inner cylinder creates a wide shear band away from the side walls and thus free of wall effects. The observed wide shear band satisfies a number of scaling laws which are quite extensive and robust \cite{unger2004shear,fenistein2004universal,ries2007shear,jop2008hydrodynamic}. Further, the tails of the velocity profile decay as an error function \cite{dijksman2009granular,ries2007shear,schall2009shear}. The flow pattern is influenced by three factors: the split position $R_\mathrm{s}$, the filling height $H$ and the rotation rate $f_\mathrm{rot}$ related to the geometry. Although the flow is purely in azimuthal direction and rate independent for small rotation frequency $f_\mathrm{rot}$, it is influenced by the flow rate at higher rotational rate due to the influence of frictional dissipation. Additionally, the flow is influenced by presence of interstitial liquid due to the effect of additional force. Thus, the flow of granular materials and the shear band formation are influenced by several independent factors. This motivates us to explore the surface velocity profile of granular materials for varying conditions by PTV-CG combined method.

\subsection{Particle Tracking Velocimetry}\label{PTV}
Particle Tracking Velocimetry (PTV) is a Lagrangian approach of measuring the velocity of individual particles by tracking their positions in several successive frames. We use two-dimensional PTV in which the surface velocity is measured. Tracer particles on the surface flow allow tracking each of them individually for several frames. The correct choice of tracer particles is critical to the successful execution of PTV experiments. The source of PTV signals is the scattering of the tracer particles, and thus the physical properties of those particles influence signal quality. Particle size, composition, density, shape, and concentration are important factors when selecting tracer particles. Tracer particles should be small enough to follow the flow being measured, but large enough to generate a strong signal. The physical properties of these particles should be close to the properties of the bulk granular particles to guarantee that they properly represent the flow behaviour. Thus, we use identical black tracer particles that are added as resident tracers to the bulk. The two-dimensional (2D) PTV, in which the flow field is measured at the surface, requires a low density of the tracer particles to allow for tracking each of them individually for several successive frames. Since the shutter speed of the camera allows a very short exposure time, the illumination of the system has to be strong and homogeneous enough for the cameras to see the light reflected by the granular particles, including the tracer particles, in every part of the measurement field. 
\par
The data images were individually digitized and stored as ($1120 \times 744$) pixels by a CCD camera oriented vertically over the illuminated surface, fixed using a tripod. The particles of interest should be ideally between $5$ to $20$ pixels in diameter. In our present study, the mean diameter of particles is equivalent to $8$ pixels. The test section was illuminated by LED light bands and recorded at the rate of $120$ frames per second as a standard. The shear cell is allowed to run initially for a given time interval $t_i = 2/f_\mathrm{rot}$ (two complete rotations) until a steady state is achieved. We ensured that a steady state is achieved when the velocity profiles of consecutive time windows overlap within an error of $1\%$. The camera is turned on thereafter to capture frames for analysing surface velocity profile in the steady state. At least $3000$ images are taken to get sufficient images for the measurement.
\par
\begin{figure*}[!htb]
  \begin{center}
    \subfigure[~Original image from Experiment \textbf{Rf1}, $f_\mathrm{rot} = 0.07$ s$^{-1}$.]{\label{fig:edge-a}\includegraphics[scale=0.22]{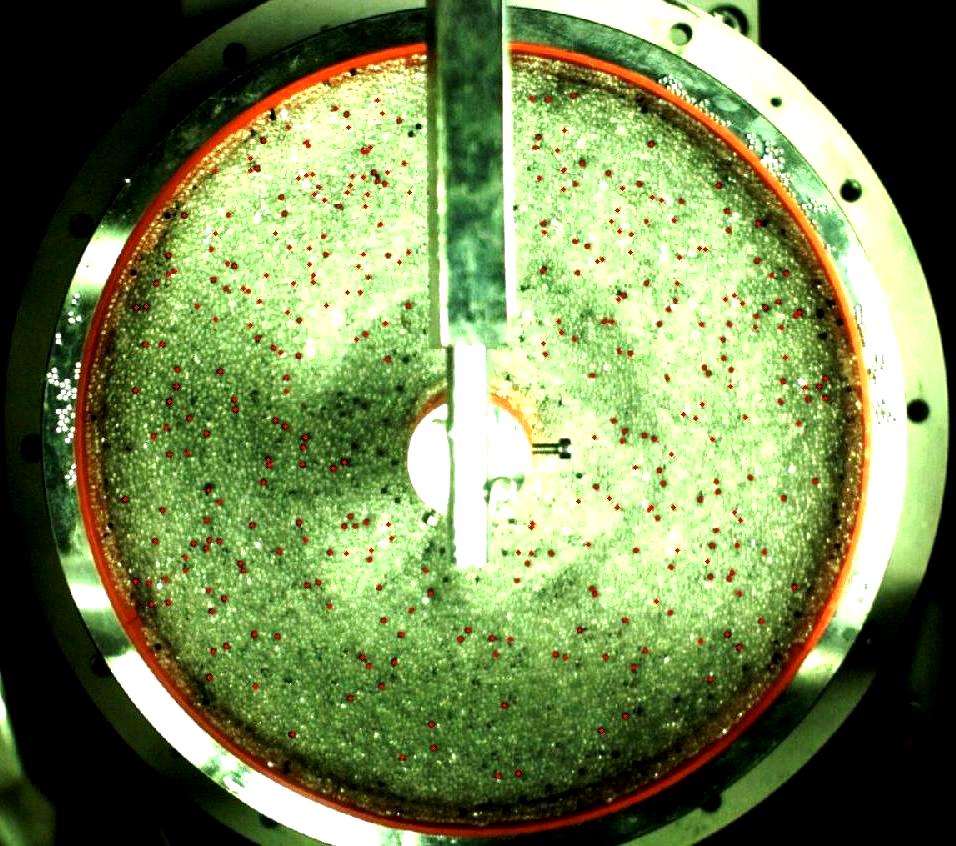}} 
    \hspace{5mm}
    \subfigure[~Inverted gray scale image from (a); geometric features have been matched.]{\label{fig:edge-b}\includegraphics[scale=0.58]{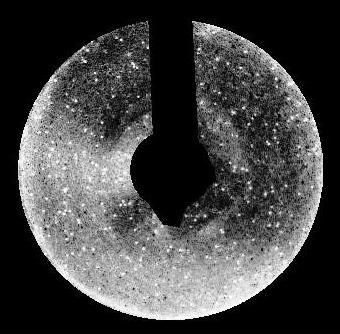}} \\
    \subfigure[~Filtered grayscale image, converted from (b) with red dots representing the tracked particles.]{\label{fig:edge-c}\includegraphics[scale=0.32]{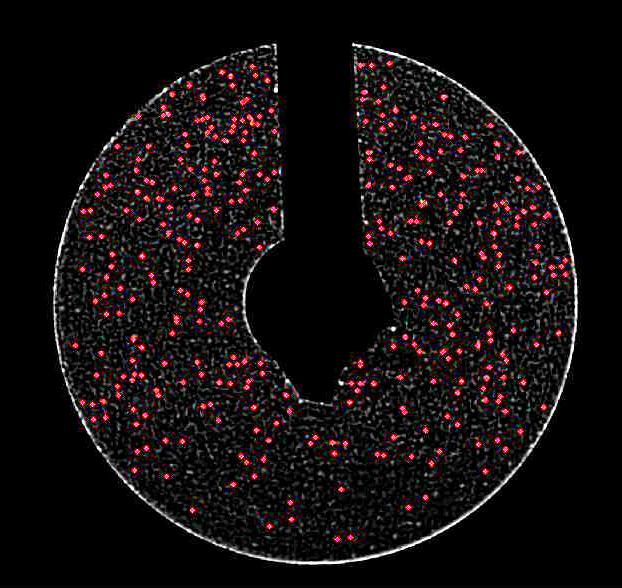}} 
    \hspace{1mm}
    \subfigure[~Particle positions tracked over multiple frames, with colours indicating differences between snapshots with increasing time from blue (first appearance of tracked particle) to yellow ($N$th appearance of tracked particle, here $N = 3000$ is the last image tracked).]{\label{fig:edge-d}\includegraphics[scale=0.1]{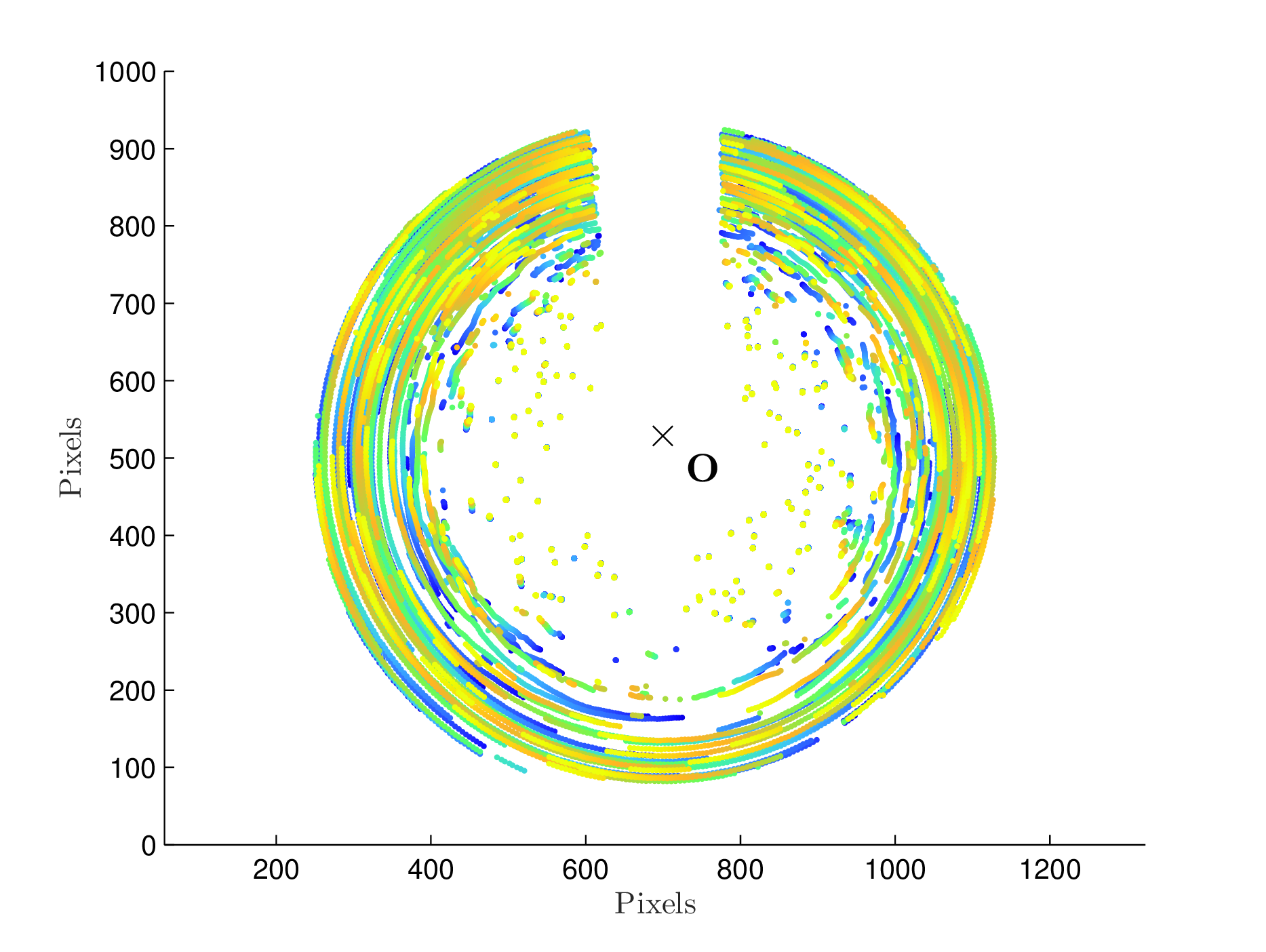}} 
  \end{center}
  \caption{Steps of 2D Particle Tracking Velocimetry (PTV) shown for images taken from Experiment \textbf{Rf1}, $f_\mathrm{rot} = 0.07$ s$^{-1}$.}
  \label{fig:edge}
\end{figure*}

We use a particle tracking code written in Matlab by adapting the \textit{IDL Particle Tracking} software, originally developed by Crocker and Grier \cite{crocker1996methods} for colloidal particle tracking. This Particle Tracking code was more generalised by Blair and Dufresne \cite{PTV} for applications in other related fields. We demonstrate here the steps for processing the captured images in Figure \ref{fig:edge}. The main features of the particle tracking algorithm are image inversion, \textit{bandpass filtering}, particle detection and linking of the particle locations from the trajectories. First, we do an inversion of the image to improve the contrast as shown in Figure \ref{fig:edge-b}. The background is subtracted from the image by masking (a non-destructive process of image editing) the area. Next, we detect the particles by finding the peaks at the centroids of the tracer particles on the surface. In this step, we also make the inessential regions of the figure to turn black. Followed by this, we do a spatial filtering of the image by using a bandpass filter. This filtering not only enhances the edges by suppressing the low frequencies, but also reduces the noise by attenuating the high frequencies. Thus, we locate the sharp peaks as the particles represented by the red dots in Figure \ref{fig:edge-c}. Likewise, we find the surface particle locations for all the images from different time snapshots. The tracker finds the shift of each particle between two consecutive snapshots. Thus, we get the trajectories of individual tracer particles over time by tracking the tracer particles over all snapshots as shown in Figure \ref{fig:edge-d}. Different colours in the figure represent increasing snapshots from blue to yellow colour. Finally, the velocity vectors corresponding to the displacement between matching particles in the snapshots are determined from the corresponding shift and time differences between the snapshots (figure not shown).

\subsection{Coarse-graining: Discrete to continuum velocity field}\label{coarse}
We use the coarse graining toolbox MercuryCG to translate the discrete particle velocities to a continuous velocity field $u$, averaged over time in a $100$ by $100$ grid spatially resolved fields \cite{weinhart2012discrete,tunuguntla2016discrete}. From statistical mechanics, the microscopic density of tracer particle $i$ in our PTV data for a system at point $r$ and time $t$ is given as:
\begin{equation} \label{rho_micro}
\rho^{mic} = \sum{m_i\delta(r-r_i(t))},
\end{equation}
where, $\delta(r)$ is the Dirac delta function, $r_i$ is the position and $m_i$ is the mass of the tracer particle $i$. To extract the macroscopic density field $\rho(r,t)$ of the tracer particles, the microscopic tracer density given by Eq. (\ref{rho_micro}) is convoluted with a spatial coarse-graining function $\psi(r)$, e.g. a Heaviside, Gaussian or a class of Lucy polynomials, thus leading to:
\begin{equation} \label{rho_macro}
\rho(r,t) = \sum{m_i\psi(r-r_i(t))} = \sum{m_i\psi_i},
\end{equation}
where $\psi(r)$ is also known as a smoothing function. For simplicity, seen later, we define $\psi_i = \psi(r-r_i(t))$.
\par
\begin{figure}[!htb]
  \begin{center}
    \includegraphics[width=0.9\columnwidth]{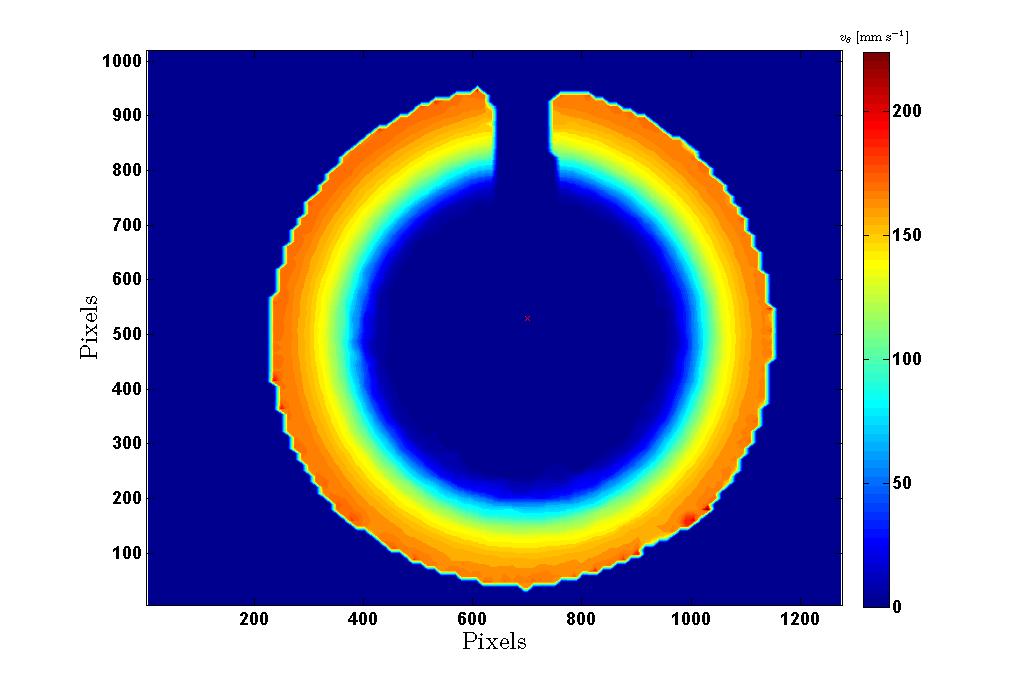}
  \end{center}
  \caption{Coarse-grained velocity field as obtained by processing data from experiment \textbf{Rf1}, $f_\mathrm{rot} = 0.07$ s$^{-1}$ using MercuryCG toolbox.}
  \label{fig:CG}
\end{figure}

Using the same idea as explained in the previous paragraph, the momentum $\textbf{P}(r,t)$ is obtained as:
\begin{equation} \label{momentum_macro}
\textbf{P}(r,t) = \sum{m_i{v_i}\psi_i},
\end{equation}
Next, we define the bulk velocity as:
\begin{equation} \label{vel_macro}
\textbf{u}(r,t) = \textbf{P}/\rho.
\end{equation}
The above definition of velocity satisfies mass conservation for any smoothing function $\psi(r)$.
We use a Gaussian smoothing function with a coarse graining width \cite{goldhirsch2010stress} equal to the mean particle diameter $d_\mathrm{p}$. In this way, we translate the discrete velocity $v_i$ data for each particle $i$ into a continuous velocity field $u$ as shown in Figure \ref{fig:CG}. Followed by this, we obtain the tangential component of the velocity field $u_\theta$ in each local grid. Low particle density in certain regions result in a CG-error and thus erroneous velocity field. Therefore, we eliminate the data corresponding to a coarse grained density $\rho$ less than $20\%$ of the mean density field from our results. 
The rotational velocity of the outer cylinder $U_{\theta} = 2\pi{R_\mathrm{o}}f$ is tracked by the PTV by tracking a fixed location on the outer rotating rim of the shear cell. The surface flows satisfy a set of scaling laws. Denoting the ratio between the observed azimuthal velocity $u_\theta$ normalised by the imposed external rotational velocity $U_{\theta}$ as $\omega = u_\theta/U_{\theta}$, and the radial coordinate as $r$, we can capture the flow profiles of shallow layers ($H/R_\mathrm{s} < 0.45$) by:
\begin{equation} \label{erf}
\omega(r) \approx \omega_\mathrm{fit}(r) = \frac{1}{2} + \frac{1}{2}\mathrm{erf}\bigg(\frac{r-R_\mathrm{c}}{W}\bigg)
\end{equation}
where, $R_\mathrm{c}$ denotes the position of the shear band centre and $W$ denotes its width on the free surface. Thus, we get an estimation of the width and location of each shear band from the surface velocity profiles.

\subsection{Locating the centre and radius}
\begin{figure}[!htb]
  \begin{center}
    \includegraphics[width=0.8\columnwidth]{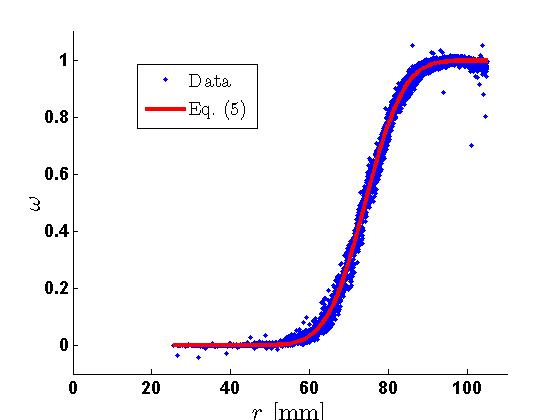}
  \end{center}
  \caption{Velocity profile as a function of radial position for a filling height of $H = 36$ mm for experiment \textbf{Rf1}, $f_\mathrm{rot} = 0.07$ s$^{-1}$. The blue dots correspond to the coarse-grained data for the experiment \textbf{Rf1}. The red solid line represents the fitting given by Eq. (\ref{erf}).}
  \label{fig:vel_profile}
\end{figure}

Locating the centre and radius of the shear cell on the PTV images of particles is essential for determining the accurate velocity profile $\omega$. Only an accurate determination of the centre gives a good collapse of the data points, which collapse well onto a curve that can be fitted by Eq. (\ref{erf}), as shown in Figure \ref{fig:vel_profile}. 
\par
There are several methods discussed here for determining the centre and the outer radius of the shear cell. First, we discuss the method of determining the centre from the discrete position and velocity data obtained by particle tracking. In order to locate the centre $\vec{r}_o$, we assume that the average particle velocity should be orthogonal to the radial direction of the particle, i.e. $\vec{v}_i \perp (\vec{r}_i - \vec{r}_o)$. Thus, we need to find the $\vec{r}_o$ that minimises the radial velocity components $v_i^r = (\vec{r}_i - \vec{r}_o)\cdot\vec{v}_i$. Thus, the sum of the squared residuals to be optimized is given by:
\begin{equation} \label{least_sq}
S_\mathrm{res} = \sum{{[(\vec{r}_i - \vec{r}_o)\cdot\vec{v}_i]}^2}
\end{equation}
The above equation is minimised using least squares \cite{umbach2003few,bevington1993data,chang2007constrained} to locate the centre labeled as \textbf{O} as shown in Figure \ref{fig:edge-d}. This is a novel application for locating the centre of a circle by fitting the position and velocity vector of the particles. The stationary particles do not contribute to the velocity and the particles in the shear band region does not have a regular velocity in the tangential direction. Thus, these particles are eliminated and only the moving particles are considered for obtaining the centre by the above mentioned method.
\par

The centre of the circumference covered by the particles in the shear cell was also determined alternatively from the coarse-grained continnum fields. This is obtained by least square minimisation of the residual in Eq. (\ref{least_sq}) and by the same methodology as explained before. The difference in approximation of the centre from the discrete and the continuum data set is within $1$\% relative to the centre obtained from the discrete data. 
\par
\begin{figure}[!htb]
  \begin{center}
  {%
  \includegraphics[width=0.9\columnwidth]{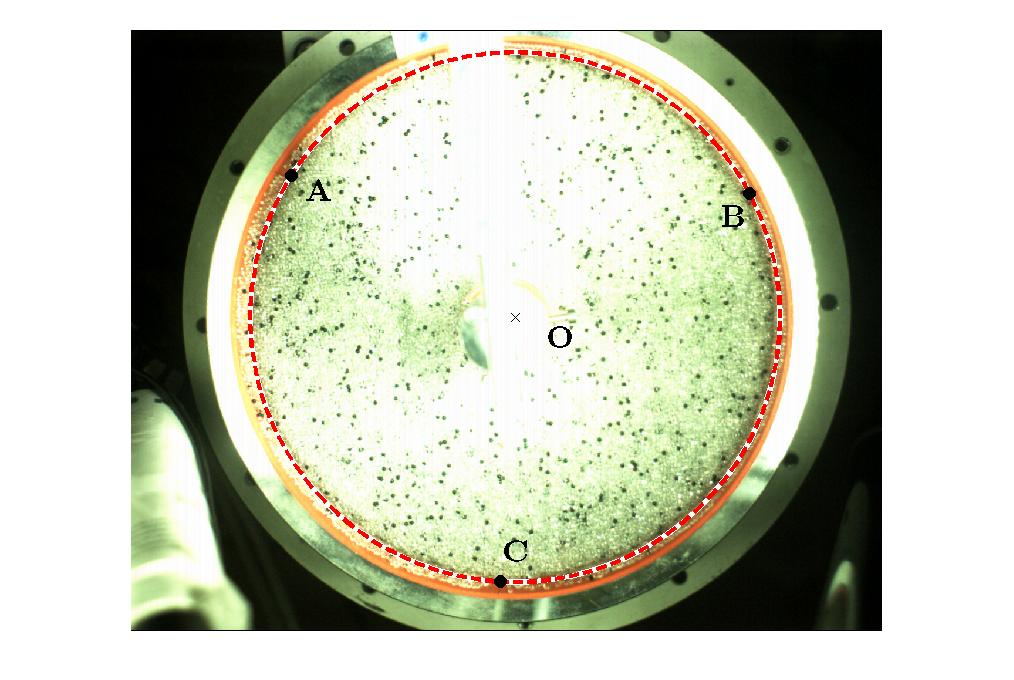}
  }%
  \end{center}
  \caption{Locating the centre \textbf{O} and radius of the shear cell from the three points \textbf{A}, \textbf{B} and \textbf{C} lying on the inner circumference of the shear cell.}
  \label{viscircle}
\end{figure}

A third way of determining both the centre and the radius is by locating three points \textbf{A}, \textbf{B} and \textbf{C} which lie exactly on the circumference as represented in Figure \ref{viscircle}. The centre and radius of the circle passing through these three points are obtained. The centre is marked as \textbf{O} in the figure. The difference in approximation of the centre by this method is also within $1$\% relative to the centre obtained from the discrete data. Thus, all three methods are equally good for obtaining the centre location. We use the last mentioned method for determining the centre of the shear cell.
\par
The radius is obtained in terms of pixel from the image analysis. This radius in pixels is equivalent to the geometric inner radius of the outer section of the shear cell ${R_\mathrm{o}}^\mathrm{in}$. Thus, we scale the radius to a general length scale in mm. A typical scaling ratio for a filling height of $36$ mm is $7.7$ pixels per mm. However, this ratio changes with the distance of the object from the location of the camera and is measured for each data set as a part of post-processing. Figure \ref{fig:vel_profile} shows a typical velocity profile at the free surface as a function of the radial position which is fitted by Eq. (\ref{erf}) to obtain $W$ and $R_\mathrm{c}$. 
\subsection{Locating the split position}
The most general way of measuring the split radius of the inner circle (${R_\mathrm{s}}^\mathrm{in}$) and the outer circle (${R_\mathrm{s}}^\mathrm{out}$), respectively, is by measuring the radius of the red insert shown in Figure \ref{Wall} by a ruler. The radius of the inner red insert is ${R_\mathrm{s}}^\mathrm{in} = 82.5$ mm, the outer insert split radius is ${R_\mathrm{s}}^\mathrm{out} = 83.5$ mm and the slit between the inner and the outer frame varies within $1$ mm. The mean split radius is labeled as $R_\mathrm{s} = 83$ mm.
\par
\begin{figure}[!htb]
  \begin{center}
    
        \includegraphics[width=0.8\columnwidth]{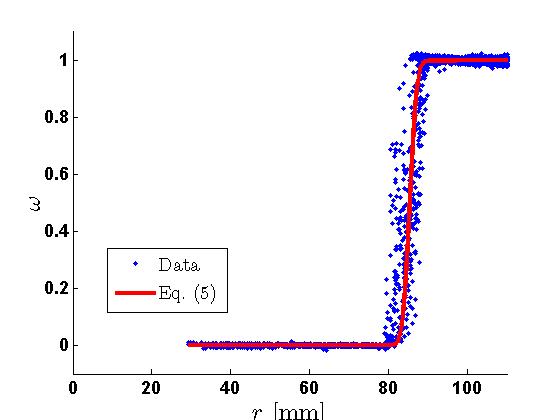}

  \end{center}
  \caption{Velocity profile as a function of radial position for the reference empty shear cell to locate the split position. The blue dots correspond to the coarse-grained data for the experiment done in empty shear cell. }
  \label{reference}
\end{figure}

An alternative method of measuring the split radius in case of a rough walls is by particle tracking in an empty rotating shear cell. The particles glued on the bottom surface  as shown in Figure \ref{Wall}, are tracked by PTV and thus the velocity profile is obtained. The split position demarcates the stationary region from the rotating outer cylinder. The velocity profile as a function of the radial position in this case is a $\delta$ function as shown in Figure \ref{reference}. Presumably, $W = 0$ mm for the data corresponding to an empty shear cell, $H = 0$ mm. However, a more accurate estimation of the split position $R_\mathrm{s}$ is obtained by fitting the data with Eq. (\ref{erf}) and a very small finite width $W \approx 2$ mm. The split position is obtained at $R_\mathrm{s} \approx 83$ mm as indicated in Figure \ref{reference}.
\par
Both the above methods of locating the split radius give approximately the same result with $R_s = 83$ mm. Thus, we consider $R_s = 83$ mm for the plots shown in Sec. \ref{dry} and \ref{wet}.

\section{Experiments with dry glass beads}\label{dry}
In the following subsections, we analyse and discuss the effects of varying filling height and shear rate on the width and position of the shear band on the free surface for dry granular materials. 

\subsection{Varying filling height}\label{Cases}
For increasing filling heights, the shear bands grow wider well, until $H < 0.45R_s$ and evolve towards the inner cylinder according to a
simple, particle-independent scaling law \cite{fenistein2004universal,dijksman2009granular,fenistein2003kinematics}. As a preliminary work and to verify with the existing established theory, we do experiments with varying filling heights $H$ for both the cases of smooth (\textbf{S}) and rough (\textbf{R}) wall surface to measure the surface flow properties for dry granular materials in the following subsections. Thereby, we compare our results of bulk flow with smooth (\textbf{S}) and rough (\textbf{R}) walls respectively for dry glass beads flow. These are essentially also prerequisite for the experiments with wet glass beads. 
\begin{table*}[htb!]
\caption{Results for two independent experiments ($1$ and $2$) with dry glass beads for varying filling heights and $f_\mathrm{rot} = 0.03$ s$^{-1}$.}
\label{appH}       
\begin{tabular}{lllllllll}
\hline\noalign{\smallskip}
 \textbf{Smooth Walls} & \textbf{Filling height} $H$ [mm] & 13 & 18 & 23 & 28 & 32 & 38\\ 
 \hline\noalign{\smallskip}
\multirow{2}{*}{$\textbf{Sh1}$} & Width $W$ [mm] & 6.45 & 6.78 & 7.49 & 7.24 & 7.23 & 8.67\\ 
 & Position $R_\mathrm{c}$ [mm] & 80.10 & 76.92 & 73.48 & 72.29 & 68.39 & 68.84\\ 
 \hline\noalign{\smallskip}
\multirow{2}{*}{$\textbf{Sh2}$} & Width $W$ [mm] & 6.69 & 6.54 & 7.62 & 7.21 & 7.31 & 7.87\\ 
 & Position $R_\mathrm{c}$ [mm] & 80.23 & 76.67 & 73.58 & 72.14 & 68.37 & 68.21\\ 
\noalign{\smallskip}\hline
\end{tabular}
\begin{tabular}{lllllllllllllll}
\hline\noalign{\smallskip}
 \textbf{Rough Walls} & \textbf{Filling height} $H$ [mm] & 6 & 8 & 10 & 13 & 15 & 18 & 20 & 23 & 25 & 28 & 32 & 38\\ 
 \hline\noalign{\smallskip}
\multirow{2}{*}{$\textbf{Rh1}$} & Width $W$ [mm] & 4.70 & 5.46 & 5.60 & 6.16 & 6.30 & 7.25 & 7.52 & 8.27 & 8.43 & 8.75 & 9.95 & 11.07\\ 
 & Position $R_\mathrm{c}$ [mm] & 83.90 & 84.19 & 84.24 & 82.57 & 82.15 & 82.29 & 81.55 & 81.04 & 80.33 & 79.17 & 78.44 & 75.23\\ 
 \hline\noalign{\smallskip}
\multirow{2}{*}{$\textbf{Rh2}$} & Width $W$ [mm] & 5.03 & 5.04 & 5.62 & 6.12 & 6.46 & 7.18 & 7.51 & 8.14 & 8.40 & 8.93 & 9.77 & 11.22\\ 
 & Position $R_\mathrm{c}$ [mm] & 83.84 & 83.8 & 84.44 & 82.67 & 82.43 & 82.28 & 81.83 & 81.4 & 80.38 & 79.57 & 78.44 & 75.66\\ 
\noalign{\smallskip}\hline
\end{tabular}
\end{table*}

For shallow layers (low filling heights), a narrow shear band develops above the split position with centre $R_\mathrm{c} < R_\mathrm{s}$. When $H$ is increased, the shear band widens continuously, until $H$ reaches $H^*$. However, the centre of the shear band $R_c$ is independent of the grain shape, size and properties \cite{unger2004shear}. Therefore, the only relevant length scales for $R_\mathrm{c}$ are the geometric scales $H$ and $R_\mathrm{s}$. The dimensionless displacement of the shear band, $(R_\mathrm{s} - R_\mathrm{c})/R_\mathrm{s}$, should thus be a function of the dimensionless height $(H/R_\mathrm{s})$ only and the power law relation for the distance from the split to the shear band centre
\begin{equation} \label{position}
\frac{R_\mathrm{s} - R_\mathrm{c}}{R_\mathrm{s}} = a{\bigg(\frac{H}{R_\mathrm{s}}\bigg)}^{5/2},
\end{equation}
 fits the data well, where $a = 0.70$. This fit to our experiments confirm other experimental and simulation findings that the shear band centre moves inwards with $R_c \propto H^{5/2}$ \cite{luding2008effect,unger2004shear,cheng2006threedimensional,torok2007shear}.
\par
The relevant length scale for the shear band width $W$ defined by Eq. (\ref{width}) is given by the grain size, and is independent of $R_\mathrm{s}$ \cite{fenistein2004universal}. First of all, $W$ grows with $H$ and increases for larger particles. $W$ grows faster than $\sqrt{H}$ as diffusion would suggest, but slower than $H$. Note that grain shape and type also influence $W(H)$: irregular particles display smaller shear bands than spherical ones of similar diameter \cite{fenistein2004universal}. The best available experimental data show that the width of the shear band is related to the filling height $H$ and particle diameter $d$ as:
\begin{equation} \label{width}
\frac{W}{d_\mathrm{p}} = b{\bigg(\frac{H}{d_\mathrm{p}}\bigg)}^{2/3},
\end{equation}
where $b = 0.84$.
 Discrete simulations, continuum models and experiments confirm these relations \cite{cheng2006threedimensional,depken2007stresses,luding2008effect,ries2007shear,henann2013predictive}. 

\begin{figure}[!htb]
  
	\includegraphics[width=0.8\columnwidth]{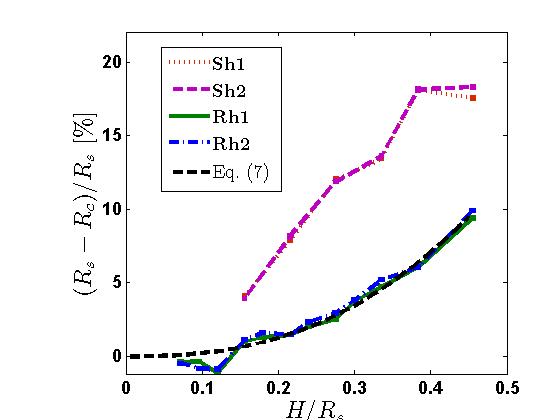}
	\hfill
\includegraphics[width=0.8\columnwidth]{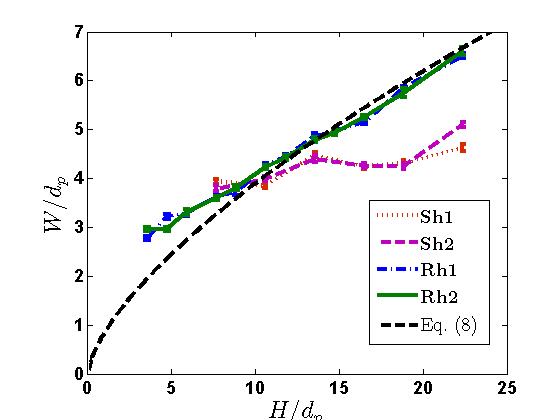}
  \caption{(a) Relative shift of the shear band centre $(R_\mathrm{s}-R_\mathrm{c})/R_\mathrm{s}$ as a function of scaled filling height $H/R_\mathrm{s}$ and (b) scaled width $W/d_\mathrm{p}$ of the shear band as a function of scaled filling height $H/d_\mathrm{p}$. The theoretical predictions of Figures (a) and (b) are given by Eqs. (\ref{position}) and (\ref{width}) with proportionality constants $a = 0.70$ and $b = 0.84$, respectively. Our experimental data in the above figures are given in Table \ref{appH}.}
    \label{fillingheight}
\end{figure}
 \par
We do experiments for measuring the location $R_\mathrm{c}$ and width $W$ of the shear band on the flow surface for shallow granular flow, varying the filling height between $6$ to $38$ mm for smooth and rough wall conditions. Figures \ref{fillingheight}(a) and (b) show a comparison of the two cases for the position of the shear band centre $R_\mathrm{c}$ and the width $W$ for varying filling heights. The dashed lines in the figures correspond to Eqs. (\ref{position}) and (\ref{width}), respectively. The points show our experimental results for smooth and rough walls. The results are also presented in Table \ref{appH}. 
\par
$R_\mathrm{c}$ and $W$ values for the experiments with rough walls agree well with the established theoretical predictions. The prefactors used to match the data in Figures \ref{fillingheight} (a) and (b) with the predictions given in Eqs. (\ref{position}) and (\ref{width}) are $a = 0.70$ and $b = 0.84$, respectively. Small deviations from the theoretical lines are observed for both $R_\mathrm{c}$ and $W$ for the experiments with very shallow layer flows. The uncertainties of data for $R_\mathrm{c}$ and $W$ are less than $1\%$ and $3\%$, respectively. However, this is expected since the distribution of the particles are not uniform over the surface and as a result the systematic contrast on the surface is not homogeneous. Moreover, there are less number of particles contributing to the experiments due to their inhomogeneous distribution over the surface which leads to a CG-error. 
\par
The experimental results with smooth wall surfaces however are not conforming to the trends of above mentioned Eqs. (\ref{position}) and (\ref{width}) as shown in Figures \ref{fillingheight}(a) and (b). For this case, we coarse-grain over different sections of the geometry to confirm that there is no effect of geometry on our results (not shown here). We obtain similar results of $R_\mathrm{c}$ and $W$ as a function of $H$ in this case from different sections of the geometry (not shown here). Nevertheless, it is evident that the shear band moves inwards with increasing filling height and also becomes slightly wider for smooth walls. 
\par
As Eq. (\ref{erf}) indicates, the flow rate away from the shear band centre decays like an error function. This is a distinguishing characteristic of wide shear band as compared with the tails of the narrow shear band, which decays exponentially. The details of the difference in the functional form of the tails of wide shear bands and narrow shear bands is discussed in \cite{schall2009shear}, which shows surface flow profiles in a Couette split-bottom cell. The resulting flows shown in this paper are similar to the split-bottom geometry for small $H$. For increasing filling heights, the flow profiles reach the inner cylinder, become independent of $H$, and exhibit exponential tails, independent of grain shape \cite{fenistein2003kinematics}. This deviation is observed for filling heights larger than $H \approx 50$ mm (data not shown), as the shear band reaches the inner cylinder.

\subsection{Varying shear rate}\label{rot}
\begin{table*}[htb!]
\caption{Results for experiments with dry glass beads for different rotation frequencies and $H = 36$ mm.}
\label{appF}       
\begin{tabular}{lllllllll}
\hline\noalign{\smallskip}
 \textbf{Smooth Walls} & \textbf{Rotation frequency} $f_\mathrm{rot}$ [s$^{-1}$] & 0.01 & 0.03 & 0.07 & 0.19 & 0.50\\ 
\hline\noalign{\smallskip}
\multirow{2}{*}{$\textbf{Sf}$} & Width $W$ [mm] & 7.71 & 7.91 & 8.17 & 8.79 & 8.98\\ 
& Position $R_\mathrm{c}$ [mm] & 67.83 & 66.74 & 66.63 & 66.61 & 67.6\\ 
\noalign{\smallskip}\hline
\end{tabular}
\begin{tabular}{lllllllll}
\hline\noalign{\smallskip}
 \textbf{Rough Walls} & \textbf{Rotation frequency} $f_\mathrm{rot}$ [s$^{-1}$] & 0.01 & 0.03 & 0.07 & 0.19 & 0.50\\ 
\hline\noalign{\smallskip}
\multirow{2}{*}{$\textbf{Rf1}$} & Width $W$ [mm] & 10.52 & 10.27 & 10.86 & 10.90 & 11.07\\ 
& Position $R_\mathrm{c}$ [mm] & 74.34 & 74.15 & 74.13 & 73.71 & 73.97\\ 
\noalign{\smallskip}\hline
\multirow{2}{*}{$\textbf{Rf2}$} & Width $W$ [mm] & 10.96 & 9.80 & 10.80 & 10.68 & 10.75\\ 
& Position $R_\mathrm{c}$ [mm] & 74.26 & 74.24 & 74.17 & 73.57 & 73.95\\ 
\noalign{\smallskip}\hline
\end{tabular}
\end{table*}
Shear localisation is a generic feature of flows in yield-stress fluids and soft glassy materials but is not completely understood. In the classical picture of yield stress fluids, shear banding happens due to stress heterogeneity. The shear band is given by a $6$ to $8$ grain diameters thick layer where the frictional dissipation is more intense than on average \cite{alonso2005micromechanics,alonso2005role}. A higher shear rate would result in a more intense frictional dissipation inside the shear band. A natural effort to minimise this dissipation would result in shifting the shear band inwards for higher shear rate \cite{unger2004shear}. Dijksman \cite{dijksman2009granular} investigated the effect of shear rate on the surface width and position of the shear band. We vary the rotation rates between  $f_\mathrm{rot} = 0.01$ to $0.50$ s$^{-1}$ keeping a constant filling height $H \approx 36$ mm. The surface velocity profiles are obtained for the varying rotation rates with images captured at a constant $120$ frames per second to conform with the threshold criteria for the highest rotational velocity of $0.50$ s$^{-1}$. The stationary velocity profiles are found to develop after two complete rotations (covered in a time interval $t_i = 2/f_\mathrm{rot}$) and remain stable thereafter in steady state. Thereby, we obtain the position of the shear band centre and its width from the flow profile as mentioned in Sec. \ref{coarse}. 
\par
\begin{figure}[!htb]
  
\includegraphics[width=0.8\columnwidth]{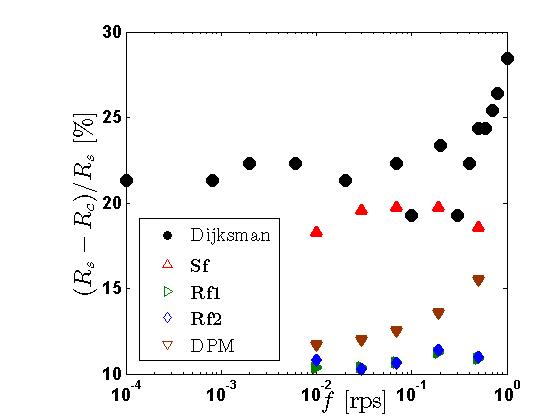}
	\hfill
\includegraphics[width=0.8\columnwidth]{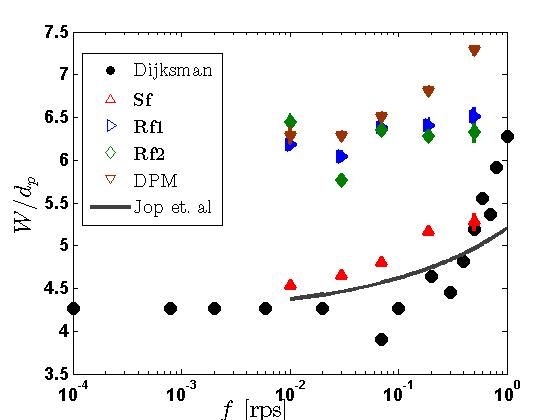}

  \caption{a) Relative shift of the shear band centre $(R_\mathrm{s} - R_\mathrm{c})/R_\mathrm{s}$ of the shear band centre b) scaled width $W/d_\mathrm{p}$ of the shear band at the surface as a function of the rotation frequency $f_\mathrm{rot}$. The solid line represents the power law $W \sim f^{0.38}$ to allow comparison with the predictions from Jop\textit{et al.} \cite{jop2008hydrodynamic} and the black solid circles $\bullet$ are the data from Dijksman \textit{et al.} \cite{dijksman2009granular}. Our experimental data in the above figures are also listed in Table \ref{appF}.}
  \label{rotational}
\end{figure}

Figures \ref{rotational}(a) and (b) show the scaled shear band centre location and the shear band width plotted against the external rotation frequency $f_\mathrm{rot}$ of the shear cell. Our experimental data are represented by the red $\triangle$. The results are also presented in Table \ref{appF}. Our experimental data is also compared with the data from \cite{dijksman2009granular}, represented by the black $\bullet$ in the figures. The shear band moves inwards (a little) and gets wider with increasing shear rate as expected.
\par
To calibrate simulation results with experiments, we ran simulations in a split-bottom shear cell geometry using the
 Discrete Particle Method in the open source code MercuryDPM. The simulations were run with a inter-particle friction $\mu_\mathrm{p} = 0.80$ and with the same particle size distribution as of the experiments, mentioned in Sec. \ref{size}. In the simulation set-up, the side walls and the bottom surface are made of particles, with roughness equivalent to the particle diameter. This represents the case of our experiments with rough walls. The data from our simulations are represented by the brown $\triangledown$ in Figures \ref{rotational}(a) and (b). The simulation results agree closely with the data from the rough wall experiments, \textbf{Rf1} and \textbf{Rf2}. However, the simulation data deviates from the experimental data at higher shear rates. The smooth experiments \textbf{Sf} data show only qualitative agreement with the Discrete Particle Method (DPM) simulations. To summarize, slower shearing does not affect the position and width of the shear band. Faster shearing moves the shear band slightly inwards and makes it wider as observed in the figure.

{\subsection{Comparison of PTV with other techniques} \label{comparison}
Among common methods for measurement of particle displacement and velocity fields in granular flows, Particle Image Velocimetry (PIV) and Particle Tracking Velocimetry (PTV) techniques are playing important roles. Their common methodological principle is that tracer particles are added to the flow, which are assumed to move with the local velocity. The surface of the flow containing the tracer particles is illuminated and recorded. Both methods determine the velocity from the displacement of particles in a moving fluid during a prescribed time interval. In contrast to PIV, in which the mean displacement of a small group of particles is sought for, PTV tracks the trajectories of individual particles along the surface across multiple images, allowing for postprocessing and thus improvement of the velocimetric data. Thus, particle identification and determination of its position in space is a very important step in PTV. This is only possible if the number density of tracer particles in an observation volume is not too high. The difference between PTV and PIV is the art of the evaluation of the recordings. In PIV, all images are subdivided into small regular sub-areas called ``interrogation areas", which are evaluated separately from each other. It is assumed that all particles within one interrogation area have moved quite homogeneously between the two recordings, and the displacements of these interrogation areas are determined by cross-correlating the interrogation area in the first frame with the two dimensional shiftings of the corresponding interrogation area. To the contrary, in PTV each individual particle is recognized and identified separately, and is resought for in the second frame. The displacement of each individual particle is effectively determined using information about the neighbourhood. The particles do not have to be homogeneously distributed in case of PTV, as it is required in case of PIV. The limit of the extent of displacements is given by the size of the interrogation area in case of PIV. In case of PTV, this limit is given by the degree of the deformations of the flow field. One of the main features and advantages of the Particle Tracking algorithm we use is that it allows us to apply a smoothing stencil for obtaining a more accurate discrete particle trajectory. Further, we use coarse graining on the discrete particle data which gives us a desired continous surface flow field consistent with the mass conservation.
\par

Besides PIV and PTV, Magnetic Resonance Imaging \cite{nakagawa1993non,ehrichs1995granular}, X-ray imaging \cite{guillard2017dynamic,baxter1989pattern}, radioactive tracers \cite{harwood1977powder}, freezing granules in resin \cite{ratkai1976particle} are also often used as other experimental methods for measuring velocity inside dense granular flow.  We discuss some of the techniques as mentioned here which were adapted by different authors for the qualitatively agreeing results shown in sections \ref{Cases} and \ref{rot}. The established scaling laws given by Eq. \eqref{position} and \eqref{width} on the shape and width of the shear band was confirmed by various authors using various techniques in the split-bottom Couette cell geometry. Fenistein  \textit{et al.} \cite{fenistein2003kinematics} established the scaling laws for the width and shape of the shear band experimentally using Particle Image Velocimetry. Cheng  \textit{et al.} \cite{cheng2006threedimensional} worked on the evolution of granular flow and established the scaling laws for the width and shape of the shear band using Magnetic Resonance Imaging technique.  Dijksman \cite{dijksman2009granular} reconfirmed these scaling laws using Particle Image Velocimetry. Our experimental data is qualitatively agreeing quite well with these scaling laws as shown in figure \ref{fillingheight}(a) and (b). Dijksman \cite{dijksman2009granular} also performed experiments for different rotation rates in shear cell geometry given by the black $\bullet$ in Figure \ref{rotational}(a) and (b) using Particle Image Velocimetry. Numerically, Luding \cite{luding2008effect} confirmed the scaling laws using the Discrete Element method and Henann \textit{et al.} \cite{henann2013predictive} confirmed Eq. \eqref{width} using the Finite Element method. Unger \textit{et al.} \cite{unger2004shear} measured the shape of the shear band as a function of the cell geometry using the principle of minimum dissipation of energy. Thus, we analyse our data using PTV technique, but we validate our results with the existing results from the literature using some of the different techniques as mentioned here.}
\section{Experiments with wet glass beads}\label{wet}
Previous studies showed that the inter-particle cohesion has a strong influence on the position and width of the shear band \cite{singh2014effect}. Such inter-particle cohesion can be introduced by adding interstitial liquid to the glass beads. The objective of this section is therefore to study the effect of interstitial liquid, glycerol, on shear band properties. Several experiments were done for different saturation of glycerol in the system as given in Table \ref{appG}. We varied the glycerol content from $0$ to $100$ ml, added to bulk material of $1.97$ kg (corresponding to a filling height of $36$ mm). This corresponds to a varying saturation of $0$ to $0.218$. 
\par
\begin{figure}[!htb]
  \begin{center}
  {%
	\includegraphics[width=0.8\columnwidth]{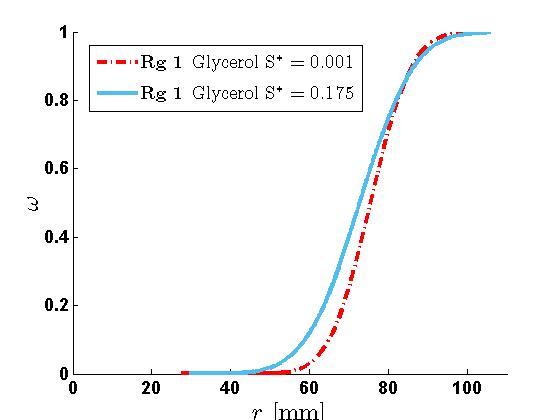}
	
   }%
  \end{center}
  \caption{Surface velocity profile $\omega$ as a function of radial position $r$ for glass beads mixed with different volumes of glycerol.}
  \label{Glycerol}
\end{figure}
\begin{figure}[!htb]
  \begin{center}
  {%
	\includegraphics[width=0.8\columnwidth]{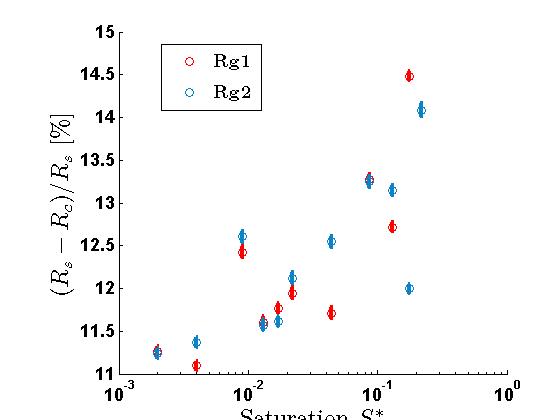}
	\hfill
	\includegraphics[width=0.8\columnwidth]{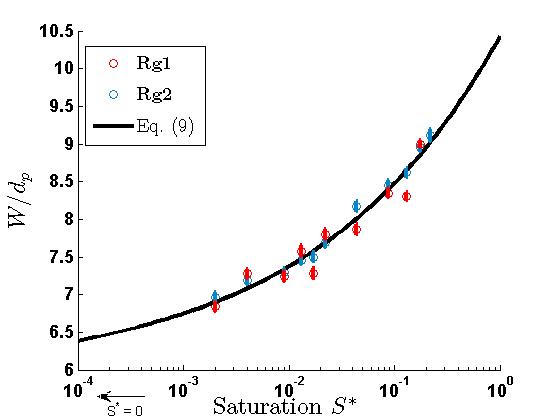}
   }%
  \end{center}
  \caption{(a) Relative shift of the shear band centre $(R_\mathrm{s} - R_\mathrm{c})/R_\mathrm{s}$ and (b) width of the shear band scaled with particle diameter $W/d_\mathrm{p}$ as a function of glycerol saturation $S^*$. The prediction represented by the solid line in Figure \ref{Glycerol-1}(b) is given by Eq. (\ref{gl_width}) with proportionality constants $W_0 = 10.07$ and ${S_0}^* = 0.002$ respectively. Our experimental data in the above figures are given in Table \ref{appG}.}
  \label{Glycerol-1}
\end{figure}

We track the velocity of the particles and apply coarse graining on discrete velocities to get the continuous velocity profile for different saturation of glycerol. Precautions are taken to prevent slippage on the inner and outer wall surfaces by making the walls rough. This allows us to get a velocity profile which can be fitted by the same error function in Eq. (\ref{erf}), including data from low to high saturation of glycerol. Note that we also did experiments of glass beads mixed with glycerol in a smooth wall surface shear cell (Experiments \textbf{Sg}) and the results are shown in \cite{roy2018hydrodynamic}. However, these results in an undesirable slippage at the inner wall of the shear cell. Figure \ref{Glycerol} shows the comparison of the position $R_c$ and the width $W$ of the shear band for different saturation $S^* = 0.002$ and $S^* = 0.175$ of glycerol. The shear band becomes wider and moves inwards with increasing saturation as shown in the figure, which is an indication of flow behaviour of cohesive materials. The results are presented in Table \ref{appG}. Figure \ref{Glycerol-1}(a) shows that the shear band centre $R_\mathrm{c}$ shifts from $11\%$ to $14.5\%$ relative to the split position $R_\mathrm{s}$. The shear band width $W$ increases in terms of particle diameter from $7d_\mathrm{p}$ at a low saturation of $S^* = 0.002$ ($1$ ml glycerol) to $9d_\mathrm{p}$ at saturation $S^* = 0.218$ ($100$ ml glycerol) as shown in Figure \ref{Glycerol-1}(b) for the same increase in saturation as in the previous figure. The uncertainties of data for $R_\mathrm{c}$ and $W$ are less than $1\%$ and $2\%$, respectively.
The data in Figure \ref{Glycerol-1}(a) data does not provide a very good fit that is convincing. However, the data in Figure \ref{Glycerol-1}(b) is fitted by a power law function given as:
\begin{equation} \label{gl_width}
\frac{W}{d_\mathrm{p}} = \frac{W_0}{d_\mathrm{p}} + {\bigg(\frac{S^*}{{S_0}^*}\bigg)}^{1/4}
\end{equation}
where $W_0$ is a fitting constant and corresponds to the width of the shear band for dry glass beads ($S^* = 0$) and ${S_0}^* = 0.002$. We expect a linear proportionality of the width of the shear band with the mean interaction distance. If the number of wet contacts are constant over increasing saturation, the volume of the liquid bridge per contact is linearly proportional to the saturation. Then the mean interaction distance is also proportional to the maximum interaction distance which is proportional to the cube root of the maximum liquid bridge volume or $S^*$ \cite{lian1993theoretical}. Thus, for constant number of contacts, the mean interaction distance is proportional to the cube root of the saturation $S^*$ in the system. However, in real case, the network of capillary bridges (number of wet contacts) increases with saturation \cite{herminghaus2005dynamics} and thus the mean interaction distance is proportional to a power, less than the cube root of the saturation $S^*$.
 \begin{table*}[htb!]
\caption{Results for two independent experiments ($1$ and $2$) with glass beads and glycerol for $f_\mathrm{rot} = 0.03$ s$^{-1}$ and $H = 36$ mm.}
\label{appG}       
\begin{tabular}{lllllllllllllll}
\hline\noalign{\smallskip}
\textbf{Glycerol saturation} $S^*$ &  & 0 & 0.002 & 0.004 & 0.009 & 0.013 & 0.017 & 0.022 & 0.044 & 0.087 & 0.131 & 0.175 & 0.218\\ 
\hline\noalign{\smallskip}
\multirow{2}{*}{$\textbf{Rg1}$} & Width $W$ [mm] & 10.27 & 11.63 & 12.37 & 12.30 & 12.88 & 12.37 & 13.25 & 13.37 & 14.19 & 14.12 & 15.30 & \\ 
 & Position $R_\mathrm{c}$ [mm] & 74.15 & 75.43 & 75.57 & 74.44 & 75.14 & 75.00 & 74.85 & 75.05 & 73.72 & 74.19 & 72.69 & \\ 
 \hline\noalign{\smallskip}
\multirow{2}{*}{$\textbf{Rg2}$} & Width $W$ [mm] & 9.80 & 11.84 & 12.22 & 12.39 & 12.68 & 12.75 & 13.07 & 13.89 & 14.37 & 14.65 & 15.21 & 15.49\\ 
 & Position $R_\mathrm{c}$ [mm] & 74.24 & 75.45 & 75.34 & 74.28 & 75.17 & 75.13 & 74.70 & 74.33 & 73.74 & 73.83 & 74.80 & 73.03\\ 
\noalign{\smallskip}\hline

\end{tabular}
\end{table*}

The shear band is the region with a strong velocity gradient and is caused by the \textit{sliding motion} of the particles. However, strong cohesive forces keep particles in contact (in other words, the cohesive forces promote \textit{collective motion} of particles) and restrict them from
sliding. As a result, the velocity gradient gets smoother and the width of the shear band increases. This observation is consistent with the previous studies on adhesive dense emulsions \cite{ovarlez2008wide} and dry cohesive dense granular flows \cite{singh2014effect}. In absence of long range forces, such an effect of cohesion is suppressed if the global Bond number is less than $1$. It has been shown that for dry cohesive materials, the global Bond number, $Bo$, captures the transition between noncohesive free-flowing granular assemblies $(Bo < 1)$ to cohesive ones $(Bo >= 1)$ \cite{singh2014effect}. However, note that in case of the dry cohesive contact model used in \cite{singh2014effect}, only closed-contact interaction forces are present. In contrast, liquid capillary bridges result in long range interactions. The Bond number for wet cohesive materials increases linearly with the surface tension of the liquid \cite{roy2016micro}. The global Bond number of glycerol on glass surface is $0.17$ corresponding to a mean pressure of $220$ Pa in the shear cell for the given filling height. Thus, the Bond number increases only slightly by adding a small amount of glycerol as compared with the dry non-cohesive materials. However, the network of capillary forces (number of wet contacts) increases significantly with increasing saturation of liquid in the pendular regime ($S^* < 0.3$) \cite{herminghaus2005dynamics}. This is due to the long range interactions between particles. Thus, we observe a significant change in the shear band features $W$ and $R_\mathrm{c}$ with increasing saturation even for $Bo < 1$ for wet granular flows which was not observed for dry cohesion \cite{singh2014effect} where the effect kicked in only at $Bo >= 1$.

\section{Conclusions and outlook}\label{con}
In the performed experiments we determined granular steady state shear band properties from the free surface velocity profile in a split-bottom shear cell. Previous studies focused primarily on dry granular flow, while here, we perform experiments on wet granular flow: glass beads with interstitial liquids as silicon oil or glycerol. This paper describes the novel approach of implementing the PTV-CG combination to obtain discrete to continuum data from experiments. The MercuryCG toolbox which is primarily designed for coarse-graining data from DPM simulations, i.e. determining of continuum fields, is proven to be applicable on discrete experimental PTV data as well.
\par
Studies on dry glass beads provide the same scaling relations as obtained earlier by various alternative methods, see subsection \ref{comparison}, which predict how the shear band moves inwards and gets wider with increasing filling height. However, there is a significant difference in the flow behaviour for smooth wall surfaces. The results of the shear band properties on the free flowing surface in a shear cell with smooth wall surfaces are not conforming to the studies done before. However, results from experiments done with varying filling heights in a shear cell with rough walls are agreeing very well with the previously established results. Particle-wall slippage is an undesirable phenomenon in our present state of smooth wall experiments. This slippage is most likely to be affected at the bottom near the split position where the shear force is highest. The shear band slips inwards towards the centre of the shear cell and thus the centre of the shear band also moves inwards. Unlike the experiments with the rough walls, here the shear band cannot grow fully to a wide shear band close to the free surface. Experiments also showed that the shear band moves inwards and gets wider with increasing rotation frequency of the shear cell. These results are in agreement with the DPM simulations at lower shear rates but deviates at higher shear rates with possible reason that we do not model all visous effects in our DPM simulations that might be more important at higher shear rates.
\par
Finally, we studied the effect of wet cohesion on the shear band properties. Experiments with glycerol added as interstitial liquid, resulted in a significant change in the shear band properties while silicone oil did not lead to much change relative to the dry case. A probable reason is, the wetting properties of the silicon oil is reduced due to its higher viscosity. The shear band gets wider and moves inward towards the inner cylinder due to the increased tendency for \textit{collective flow} of cohesive materials in case of glycerol. These results are consistent with some of the numerical studies done earlier for dry cohesive powders.
\par
Our future investigations will concentrate on the measuring the bulk behaviour of liquid migration in unsaturated granular media from image analysis and also on comparing PIV and PTV techniques in a rotating drum geometry. Some ongoing studies include theoretical analysis of transport mechanism for liquid migration in partly saturated wet granular media \cite{roy2019drift} and rheology of granular materials in different geometries \cite{shi2018steady}.
\section*{Acknowledgement}
We thank S. Zhao for helping us with the fluid-property measurements in the Physics of Fluids group, UTwente. We would also like to thank J. Dijksman for helpful advice. Finally, we thank NWO/STW for financial support of the Project 12272 ``Hydrodynamic theory of wet particle systems: Modeling, simulation and validation based on microscopic and macroscopic description''.

{\section*{Author contributions}
T.W. and A.R.T. conceived of the presented idea. B.J.S. designed and developed the in-house experimental set-up. S.R. performed the experiments, did the post-processing analysis and wrote the manuscript with support from  B.J.S. and H.P. H.P. developed and predominantly performed the analysis on particle tracking velocimetry with support from A.R.T., D.R.T. and T.W. Furthermore, A.R.T and S.L. encouraged S.R. to re-investigate the experiments with rough walls. S.L. and T.W. supervised S.R. throughout the findings of this work. }

\nocite{*}

\bibliography{apssamp}

\end{document}